# Resonantly pumped bright-triplet exciton lasing in caesium lead bromide perovskites

Guanhua Ying* [1], Tristan Farrow* [1,2,†], Atanu Jana[3], Hanbo Shao[4], Hyunsik Im[3], Vitaly Osokin[1], Seung Bin Baek[5], Mutibah Alanazi[1], Sanjit Karmakar[2], Manas Mukherjee[2,†], Youngsin Park[5,†], Robert A. Taylor[1,†]

* Co-first authors

[†] Corresponding Authors, T.F. (Tristan.Farrow@cantab.net),
M.M. (cqtmukhe@nus.edu.sg), Y.P. (ysinpark@unist.ac.kr),
R.A.T. (robert.taylor@physics.ox.ac.uk)

[1]Clarendon Laboratory, Department of Physics, University of Oxford, Parks Road, Oxford OX1 3PU, UK

[2]Centre for Quantum Technologies, National University of Singapore, Science Drive 2, Singapore 117543

[3]Division of Physics and Semiconductor, Dongguk University, Seoul 04620, Korea.

[4]State Key Laboratory of Mechanics and Control of Mechanical Structures, Nanjing University of Aeronautics and Astronautics, Nanjing, Jiangsu 210016, China

[5]School of Natural Science, Ulsan National Institute of Science and Technology, Ulsan 44919, Korea

## ABSTRACT

The surprising recent observation of highly emissive triplet-states in lead halide perovskites accounts for their orders-of-magnitude brighter optical signals and high quantum efficiencies compared to other semiconductors. This makes them attractive for future optoelectronic applications, especially in bright low-threshold nano-lasers. Whilst non-resonantly pumped lasing from all-inorganic lead-halide perovskites is now well-established as an attractive pathway to scalable low-power laser sources for nano-optoelectronics, here we showcase a resonant optical pumping scheme on a fast triplet-state in $CsPbBr_3$ nanocrystals. The scheme allows us to realize a polarized triplet-laser source that dramatically enhances the coherent signal by one order of magnitude whilst suppressing non-coherent contributions. The result is a source with highly attractive technological characteristics including a bright and polarized signal, and a high stimulated-to-spontaneous emission signal contrast that can be filtered to enhance spectral purity. The emission is generated by pumping selectively on a weakly-confined excitonic state with a Bohr radius ~10 nm in the nanocrystals. The exciton fine-structure is revealed by the energy-splitting resulting from confinement in nanocrystals with tetragonal symmetry. We use a linear polarizer to resolve two-fold non-degenerate sub-levels in the triplet exciton and use photoluminescence excitation spectroscopy to determine the energy of the state before pumping it resonantly.

**INTRODUCTION**

When it became apparent that triplet-excitons in all-inorganic perovskites are optically active[1,2], and hence do not suffer from intensity-quenching due to long-lived dark-states seen in their hybrid counterparts, the idea that they could be exploited for bright polarized coherent emission above the lasing threshold followed naturally. Here we present a scheme to resonantly pump the triplets at fluences above lasing thresholds to realize coherent emission with bright polarized signals with low-incoherent background contributions. Optical transitions from triplet states are normally spin-forbidden and result in long-lived dark states compared to emissive singlet states. This might appear to be problematic for the efficiency of optoelectronics given the three-to-one prevalence of triplets over singlets in a range of optical materials including semiconductors and lead halide perovskites[1,4]. Recent reports[1,2] overturn this view by showing that the emission intensity from fast triplets in lead halide perovskite nanocrystals (PNCs) is not only bright but accounts for why these materials can be up to $10^3$ times brighter than other semiconductors. These are encouraging results for photovoltaics, single-photon sources[2], wavelength-tunable nanolasers[5-13], non-linear[14] and spintronic devices[15] and optoelectronic applications[11]. Unlike in other semiconductors, the heavy ions in $CsPbBr_3$ lead to strong coupling between the spin and orbital angular momenta of holes ($j_h = 1/2$) and electrons (where $j_e = \pm 1/2$, since the electronic state is doubly degenerate due to the stronger spin-orbit term in the conduction band[4]) such that only the total angular momentum ($J = j_h + j_e$) is conserved. The spin-degeneracy is lifted when momenta of the electron and hole are mixed through an exchange interaction revealing the exciton's fine structure with distinct singlet ($J=0$) and triplet ($J=1$) states[1,2] (Fig. 1a). Whilst the exact mechanisms of symmetry-breaking and state energy re-ordering (where the triplet energy is pulled below the singlet) producing a Rashba-type effect remain unclear[4], recent studies confirmed that triplets in lead halide PNCs become dipole-allowed and are bright. The triplet exciton resolves into three non-degenerate sub-levels (with orthogonal linear dipoles) when quantum-confined in PNCs in the orthorhombic phase, and into two states in the tetragonal phase, but reverts to a degenerate triplet in the isomorphic (cubic) phase.

**RESULTS AND DISCUSSION**

In this work, we study colloidal PNCs (see section on Materials and Methods) in the tetragonal phase where the linear dipoles are polarized along two symmetry axes. We measure fine structure splitting by placing a linear polarizer into the optical path

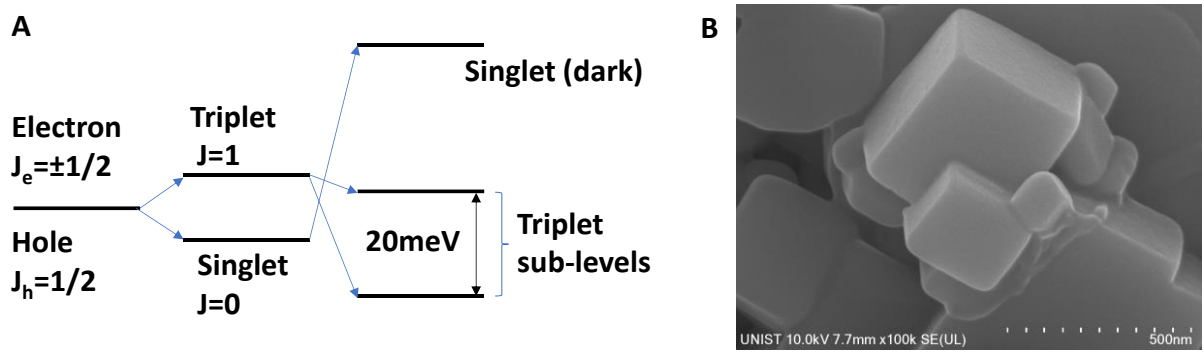

**Figure 1. (A) Excitonic energy fine structure. (B) SEM of encapsulated CsPbBr$_3$ nanocrystals**

**A.** Energy fine structure in CsPbBr$_3$. The presence of heavy ions in lead-halide perovskites leads to strong spin-orbit coupling whilst symmetry breaking produces a Rashba-type effect resulting in an inversion of the single and triplet energies and the lifting of the triplet degeneracy into dipole-allowed emissive triplet states. **B.** Scanning electron microscope (SEM) image of monolithic Cs4PbBr$_6$ perovskite microcrystals. The microcrystals encapsulate regular arrays of CsPbBr$_3$ nanocrystals with tetragonal symmetry (see Fig. S1) and have dimensions ranging from 0.1 to ~1 μm.

of the emission. The measurement is performed at cryogenic temperatures to resolve the splitting, since its energy is small compared to the thermal bath at room temperature.

The optical properties of our CsPbBr$_3$ nanocrystals are characterized via a fiber-based confocal micro-photoluminescence (μPL) setup with a tunable source for illumination. The source comprises a supercontinuum white pulsed laser operating at 78 MHz with a pulse width of 10 ps (SuperK Extreme, NKT Photonics) used together with a pair of controllable transmission gratings and an output slit to spectrally filter out a narrow band (1.5 ~ 2 nm) within the source wavelength range (Fianium fliter). The outgoing beam is then coupled into a multimode fiber for the purpose of beam-shaping before being focused through a 100x microscope objective onto a collection of microcrystals made of regular arrays of tetragonal perovskite nanocrystals with a diffraction-limited spot size of ~1.2 μm. The setup thus produces an excitation source with a spectrally-narrow, spatially-limited spot which allows the wavelength to be scanned continuously covering the full visible spectrum. The luminescence is then directed confocally to a spectrometer (spectral resolution of ~700 µeV) via a multi-mode optical fiber of 25 µm core size to limit the collection area to a spot with diameter ~1 µm at the sample. The signal is finally detected using a cooled charge coupled device (CCD) detector.

We investigated the on-resonance pumping regime of our $CsPbBr_3$ nanocrystals by tuning the excitation energy gradually into quasi-resonance with the upper-level of a triplet state (**Fig.1**(A)) at ~2.335 eV (531 nm), as determined by photoluminescence excitation spectroscopy (PLE), and a PLE spectrum is shown in **Fig S3**. **Fig. 2**(A) shows a series of emission spectra as a function of excitation wavelength with the pump laser filtered spectrally at a pump power ~50 µW, well above the lasing threshold at 4 K. A red shift of the stimulated emission (SE) peak is observed **Fig. 2**(B) as the pump wavelength tunes into resonance with the excitonic transition. Since carriers excited by a near-resonant pump occupy lower-lying energy states closer to the band-edge, it follows that upon recombination their emission wavelength is red-shifted relative to that generated by states occupying higher energy levels. The relative intensity shows the maximum near the on-resonance excitation regime (**Fig. 2**(C)). The **Fig. 2**(D) highlights the multiple dynamical processes can take place during on- /off-resonant optical excitation.

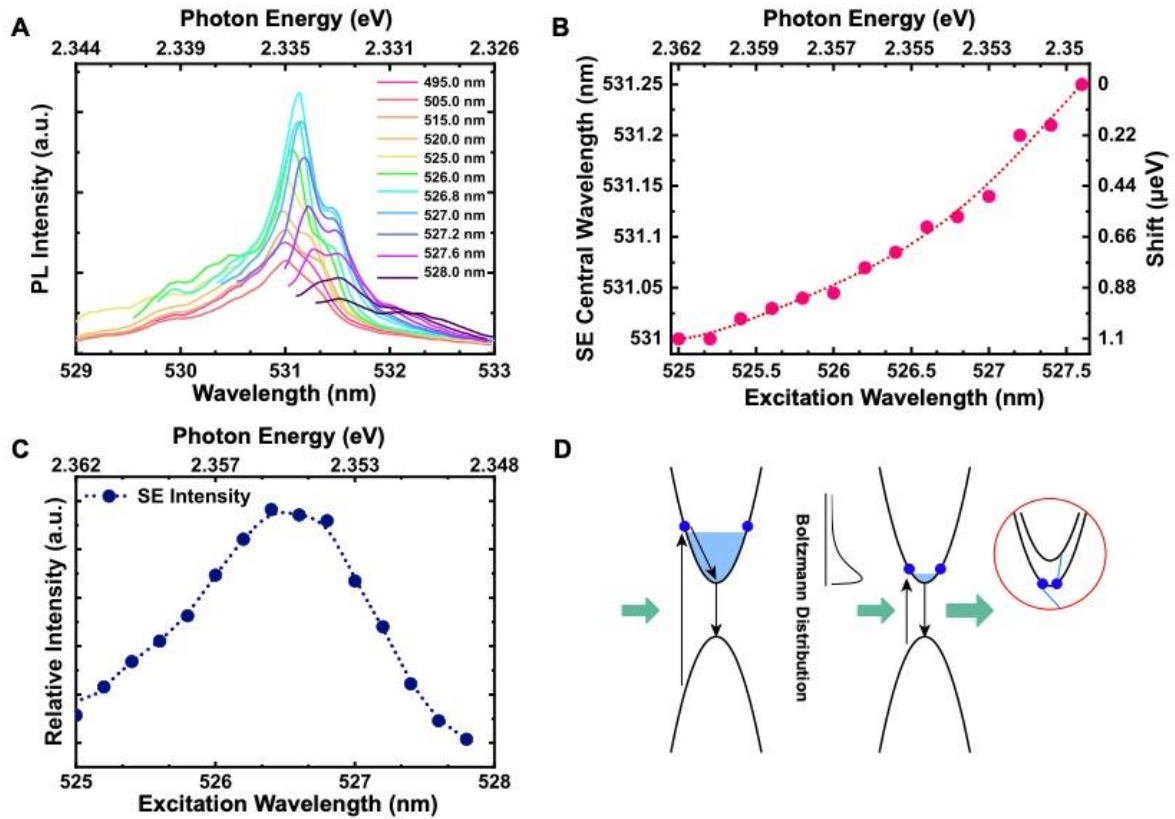

**Fig. 2 | Photoluminescence spectra from CsPbBr$_3$ PNCs as the pump wavelength is varied.**
**A.** Excitation wavelength scan using a super-continuum white laser in conjunction with an adjustable grating at 4 K. The fluence can be controlled and the scan is only performed over a narrow wavelength range where the excitation intensity is kept constant. The figures on the right are the excitation wavelengths in nm. **B.** Red shift of the SE peak after the excitation wavelength exceeds the resonant pumping wavelength with the dotted line as an eye-guide. **C.** Integrated SE intensity as a function of excitation wavelength. **D.** In non-resonant excitation of an electron-hole pair, radiative and non-radiative phonon-scattering decay pathways can generate spontaneous emission resulting in an incoherent background signal on top of the stimulated signal. For excitations near the excitonic transition, the absorption coefficient increases, resulting in brighter photoluminescence and a suppression of the incoherent decay pathways as the intensity of the coherent signal increases. The inset shows an Auger mechanism when excess carriers concentrate in a small region.

To resolve the fine structure splitting of the state we placed a linear polarizer into the optical path of the emission. A polarization-dependent analysis was then performed on the photoluminescence (PL) with linear polarizer in tandem with a half-wave plate, which was used to select a narrow polarization angle of the emission spectrum by aligning it with the spectrometer grating. The polarization-resolved PL spectra in **Fig. 3**(A) show two peaks at 2.335 eV and 2.337 eV revealing the fine-structure of the triplet sub-levels separated by 2 meV. This is consistent with measurements[1,2] for PNCs with edge lengths ~ 10 nm. Our time-resolved PL decay measurements (**Fig. S2**) yield a lifetime of 49 ps for the stimulated emission and 268 ps for the spontaneous emission, consistent with a weakly confined exciton with a Bohr radius ~10 nm[1]. We also observe a Stokes shift of 18 meV (**Fig. S3**) corresponding to the energy given up by the confined exciton to phonons in the crystal lattice combined with a broadening arising from the ensemble of nanocrystals involved in the emission and the fact that stimulated emission will always appear in regions of low optical loss. The magnitude of the effect corroborates values reported elsewhere[16] for nanocrystals of comparable size.

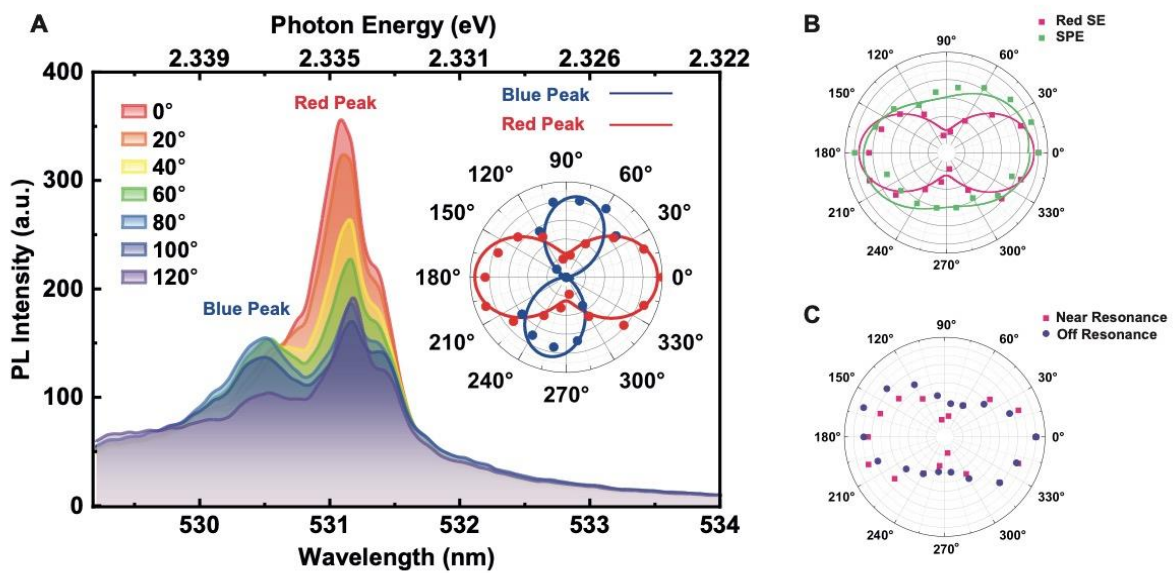

**Fig. 3 | Polarization spectra of the CsPbBr₃ nanocrystals. A.** Polarization dependent PL spectrum of the PNCs measured at 4 K with an excitation power of ~50 μW and near-resonant laser excitation. The inset depicts a polar angle diagram of the two emission peaks using a linear polarizer in tandem with a half-wave plate to select a narrow polarization angle of the emission spectrum aligned with the spectrometer gratings. Each peak's maximum intensity on the polar plot is self-normalized. **B.** A comparative study of the polarization-dependence revealed that the SE is markedly more polarized than the background signal, **C.** Polarization-dependence of the on- and off- resonance polarization signals generated with a 2.505 eV (495 nm wavelength) pump.

The near-total extinction of the emission for certain polarizer angles shown in the polar plot inset in **Fig. 3**(A) confirms that the emission is linearly polarized, and reveals the orthogonally polarized directions of two non-degenerate triplet sublevels under near-resonant excitation. The peaks are fitted with a Gaussian function (**Fig. S4**). The integrated PL intensity is fitted using Malus's law, $I(\theta) = I_{min} + I_{max} \cos^2(\theta - \theta_0)$, where $\theta$ is the of polarization angle and $I_{min}$ and $I_{max}$ are the minimum and maximum intensities, respectively. The degree of linear polarization, defined as $(I_{max} - I_{min})/(I_{max} + I_{min})$, is calculated to be over 85% for the lower energy peak and ~100 % for the higher energy peak. As the polarizer angle is rotated, the emission intensity from each state increases as the other decreases until extinction when rotated through 90°. A comparative study of the polarization-dependence (**Fig. 3**(B)) shows that the signal from the low-lying triplet state is significantly more polarized than the emission from higher energy states above the band gap. The on- /off-resonance emission associated with the red peak (**Fig. 3**(C)) shows near-total extinction when pumped resonantly, but slightly weaker intensity modulation by the polarizer when pumped above resonance with a 2.505 eV (495 nm wavelength) pump. In general, resonant excitation reduces depolarization effects and in light of the observed Stokes shift with above-resonance pumping, a contributing factor to depolarization arises from phonon scattering leading and a broadening of the confinement potential of the dipoles along the orthogonal polarization axes in the lattice. This is of technological relevance for applications where polarization-encoding of information depends on efficiently resolving the polarization basis, such as in single-photon source for quantum key distribution.

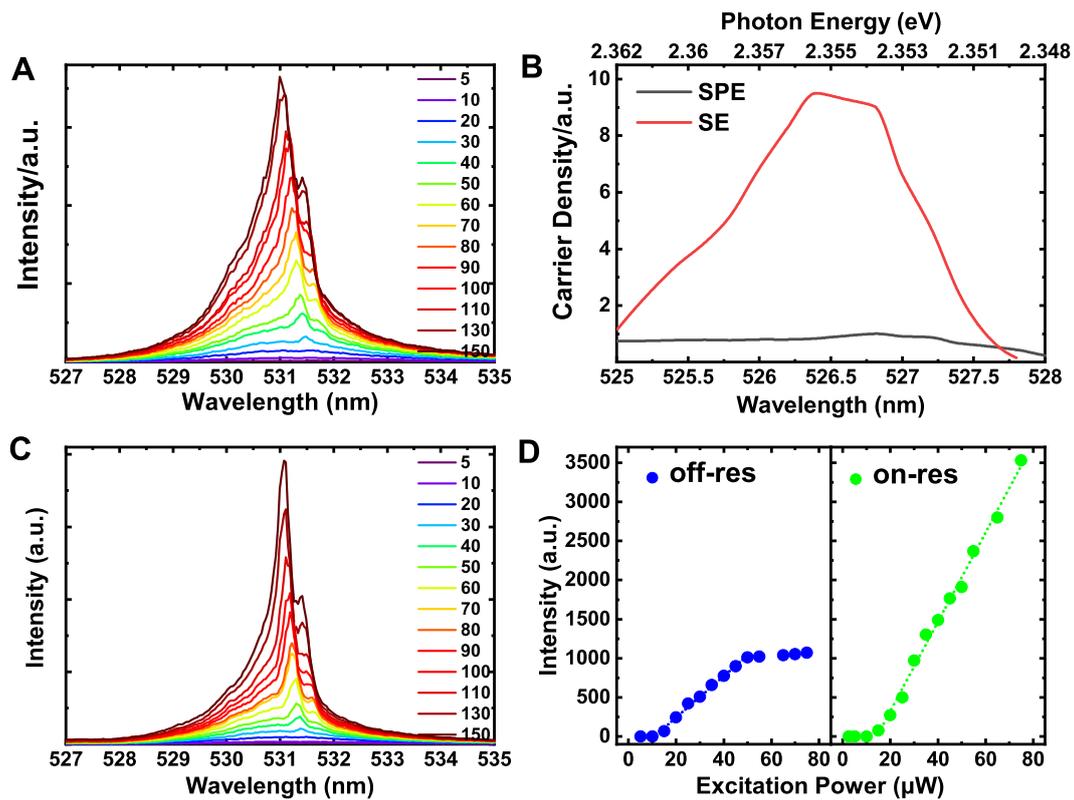

**Fig. 4 | Excitation power dependent PL spectra of the CsPbBr3 PNCs measured at 4 K.**

**A.** Intensity dependence on excitation power using an off-resonant optical pump (495 nm). The numbers relate to the excitation incident average power in µW. **B.** Nominal total carrier density compared to the carrier density contributing only to spontaneous emission (SPE) over the range of the wavelength scan. The difference between the two curves represents the integrated SE intensity. **C.** Intensity dependence on excitation power using a near-resonant optical pump (526 nm). The excitation fluence is controlled to give the same density of pumping photons. Importantly, the contrast between the SE and SPE peaks is significantly higher even at near-threshold pumping. The numbers relate to the excitation incident average power in µW. **D.** Integrated SE intensity versus pump power for non-resonant (left panel) and near- resonant (right panel) pumping plotted on the same scale for emission intensity. Over the same range of pump-photon density, the near-resonant pumping regime produced a significantly steeper linear lasing intensity increase, while saturation is not reached even at the highest pumping fluence available. In the off-resonant pumping regime, a shallower linear intensity increase is observed, reaching saturation at 50 µW.

**Fig. 4**(A) shows the µPL spectra from CsPbBr3 PNCs illuminated at increasing excitation powers with an off-resonant pulse with an excitation energy of 2.505 eV (495 nm). The PL peak around 531 nm falls within the well-known emission wavelength range of CsPbBr3 crystals at 4K[5-7]. Since the wavelength of the pump-photons is tuned above resonance, PL is generated by carriers above the bandgap

energy. Below excitation powers ~10 µW, only a broad peak with FWHM (full at half maximum) < 4 nm (18 meV) is visible, corresponding to spontaneous emission (SPE) due to incoherent recombination. The position and intensity of the peak was inferred from Gaussian fitting (**Fig. S4**). As the excitation power increases, a sharp peak emerges and dominates the spectrum. This is attributed to stimulated emission (SE), corroborated by a characteristic S-shaped curve for the lasing intensity- dependence on pump fluence (**Fig. 4**(D)) above a threshold of 15 µW (~23 mJ/cm$^2$) for lasing onset. Ultra-low thresholds have been demonstrated with PNC lasers down to 220 nJ/cm$^2$ in pulsed[17] and continuous-wave operation modes at room temperature[18], underscoring the technological viability of lead-halide perovskites as gain media for low-threshold lasers[19,20]. We note a shift of the emission wavelength to higher energy with increasing pump power. This blue-shift is due to the recombination of states with energies above the extrema of the band-edges. At high pump fluences, these states are quickly repopulated, and as the pump power increases, higher-energy states lying further away from the band edges are populated. These hot carriers thermalize by scattering with phonons and other hot carriers until they reach the crystal temperature with energies characterized by a Boltzmann distribution, resulting in the observed blue-shift of the emission[21-23].

As we approach the resonant pumping energy, the intensity of the lasing signal increases markedly relative to the incoherent signal. Conversely, we see a decrease of the lasing intensity as the pump-photon energy drops below the resonant energy, as expected. In **Fig. 4**(B) we compare the relative carrier density contributing to the SE and SPE components of the emission as a function of pump wavelength. We note a marked increase in the proportion of carriers participating in SE as we near resonance, suggesting that incoherent decay pathways are suppressed. As the pump-energy nears resonance, incident photons scatter coherently with carriers occupying the lowest energy level close to the band-edge and generate a signal via SE. In the non-resonant excitation regime, carriers excited well above the band edge can also undergo SE, albeit many will decay via fast phonon-mediated relaxation, which detunes them from resonance with the pump. Thus the fraction of carriers participating in SE is lower in the non-resonant regime. Because resonant pumping generates carriers near the band edge only, it suppresses the multiple incoherent decay pathways available to carriers in higher-energy levels. Hence it reduces incoherent contributions to the signal while enhancing lasing intensity and spectral purity. We show this explicitly in **Fig. 4**(D) and measure a tenfold enhancement the lasing intensity on-resonance, while the SPE emission intensity increases only by 30%. The fact that the on-resonance intensity continues to rise linearly with excitation power up to the maximum excitation power available at this energy (80 µW) indicates that heating effects are not important over the range of powers used and little degradation of the signal is seen for long excitation times indicating that these systems would be suitable for real-word applications.

**CONCLUSIONS**

We demonstrated that resonant pumping on a bright triplet state in $CsPbBr_3$ produces a laser source with highly-desirable technological characteristics, including a low-threshold lasing onset, a bright polarized signal, and a high SE to SPE signal contrast that can be filtered to enhance spectral purity.

**MATERIALS AND METHOD**

**Laser system and PL**

The incident laser power on the CsPbBr$_3$ surface ranged from 100 nW to a few hundred µW. The sample was mounted in a continuous-flow helium cryostat, allowing the temperature to be controlled accurately from 4 K to room temperature. Measurements are performed at cryogenic temperatures to resolve the triplet splitting, since its energy is small compared to the thermal bath at room temperature.

**Optical photoluminescence and coherent measurements**

A colloidal solution of CsPbBr$_3$ was dispersed on a quartz substrate. The optical properties of the CsPbBr$_3$ nanocrystals were characterized using a fiber-based confocal micro-photoluminescence (µPL) setup with a tunable pump source for Photoluminescence excitation (PLE) spectroscopy. We used a supercontinuum white pulsed laser operating at 78 MHz with a pulse width of 10 ps (SuperK Extreme, NKT Photonics) for the source in conjunction with a controllable transmission grating and an output slit to spectrally filter a narrow band (1.5 – 2 nm) within the pump's wavelength range. A Picoharp time-correlated single photon counting system was used in conjunction with a photomultiplier for the time-resolved measurements.

**Sample preparation**

**Reagents:** Cs$_2$CO$_3$ (99%), PbBr$_2$ (98%), Oleic acid (OA, ≥99%), 1-octadecene (ODE, tech., 90%), HBr (ACS reagent, 48%), octylamine (OAm, 99%), and diethylether (99%) were purchased from Sigma-Aldrich. Octylammonium bromide (OABr) was prepared according the previously reported method[22].

**Synthesis of colloidal CsPbBr$_3$ tetragonal nanocrystals inside Cs$_4$PbBr$_6$ microcrystals:** In a typical synthesis, Cs$_2$CO$_3$ (0.0325 g, 0.1 mmol) was dissolved in 2 ml ODE and 1 ml OA in a 15 ml glass vial under stirring condition. The solution was dried for 1 hour at 120 °C until all Cs$_2$CO$_3$ reacted with OA. Cs$_2$CO$_3$ reacts with OA to form Cs-oleate, CO$_2$, and H$_2$O. At high temperature, both CO$_2$ and H$_2$O are evaporated. The solution was kept at 150 °C to avoid solidification. In a separate vial, 2 mL of ODE, OAmBr (0.042 gm, 0.2 mmol), and PbBr$_2$ (0.073 gm, 0.2 mmol) and 1 ml DMF were heated at 120 °C in open-air. Pure OAm is detrimental to the nanocrystal surface as it may exist in dynamic equilibrium with the OA[23]. Instead of pure OAm, here we have introduced bromide ammonium salt, OABr for synthesis of the PNCs. Here 3 ml of Cs-oleate solution was injected quickly into the lead precursor solution. After cooling to room temperature in ambient conditions, the crude solution was centrifuged immediately at 5000 rpm for 5 min. Then, the supernatant was removed, and the precipitate was dried at 60 °C for further use. Generally, the CsPbBr$_3$ PNCs were synthesized using oleic acid and oleylamine, which have the same number of carbon atoms, and both the surface-passivating ligands maintain the homogeneous distribution of cubic- sized PNCs[24]. In our case, we have used OABr, which has eight carbon atoms, and the combination of short-chain OABr and long-chain OA results in a large quantity of CsPbBr$_3$ PNCs embedded regularly in Cs$_4$PbBr$_6$ microcrystals with dimensions ranging for 0.1 to ~1 μm.

**Supporting Information:** This material is available free of charge via the internet at http://pubs.acs.org. It contains TEM images of the nanocrystals, a time-resolved emission spectrum, PLE and PL plots and fitting data.

**Acknowledgements**

T.F. and M.M. acknowledge support from the Singapore Ministry of Education and the National Research Foundation, T.F. acknowledges support from the Centre for Quantum Technologies (NUS) Exploratory Fund. Y.P. acknowledges support from the National Research Foundation of Korea (NRF) grant funded by the Korea government (MSIT) (2018R1D1A1B07043676 and 2021R1A2C1006113). The Oxford authors acknowledge support from the Oxford-ShanghaiTech collaboration project.

**Competing interests**

The authors declare no competing interests.

**Data availability**

The data related to the figures are freely available at:

https://doi.org/10.5287/bodleian:E9gbwmVNk

**REFERENCES**

(1) Becker, M. A.; Vaxenburg, R.; Nedelcu, G.; Sercel, P. C.; Shabaev, A.; Mehl, M. J.; Michopoulos, J. G.; Lambrakos, S. G.; Bernstein, N.; Lyons, J. L.; Stöferle, T.; Mahrt, R. F.; Kovalenko, M. V.; Norris, D. J.; Rainò, G.; Efros, A. L. Bright triplet excitons in caesium lead halide perovskites. *Nature* **2018**, *553*, 189–193.

(2) Utzat, H.; Sun, W.; Kaplan, A. E. K.; Kreig, F.; Ginterseder, M.; Spokoyny, B.; Klein, N. D.; Shulenberger, K. E.; Perkinson, C. F.; Kovalenko, M. V.; Bawendi, M. G. Coherent single-photon emission from colloidal lead halide perovskite quantum dots. *Science* **2019**, *363*, 1068–1072.

(3) Saba, M. Rule-breaking perovskites, *Nature* **2018,** *553*, 163-164.

(4) Stranks, S. D.; Plochocka, P. The influence of the Rashba effect, *Nat. Mater.* **2018**, *17*, 381–382.

(5) Swarnkar, A.; Chulliyil, R.; Ravi, V. K.; Irfanullah, M.; Chowdhury, A.; Nag, A.; Colloidal CsPbBr$_3$ perovskite nanocrystals: Luminescence beyond traditional quantum dots.

*Angew. Chem.* **2015**, *127*, 15644–15648.

(6) Rainò, G.; Nedelcu, G.; Protesescu, L.; Bodnarchuk, M. I.; Kovalenko, M. V.; Mahrt, R. F.; Stöferle, T. Single cesium lead halide perovskite nanocrystals at low temperature: fast single-photon emission, reduced blinking, and exciton fine structure. *ACS Nano* **2016**, *10*, 2485–2490.

(7) Makarov, N. S.; Guo, S.; Isaienko, O.; Liu, W.; Robel, I.; Klimov, V. I. Spectral and dynamical properties of single excitons, biexcitons, and trions in cesium–lead-halide perovskite quantum dots. *Nano Lett.* **2016**, *16*, 2349–2362.

(8) Park, Y.; G. Ying, Jana, A.; Osokin, V.; Kocher, C. C.; Farrow, T.; Taylor, R. A.; Kim, K. S. Coarse and fine-tuning of lasing transverse electromagnetic modes in coupled all-inorganic perovskite quantum dots. *Nano Res.* **2020**, *14*, 108–113.

(9) Sutherland, B.; Sargent, E. Perovskite photonic sources. *Nat. Photonics* **2016**, *10*, 295–302.

(10) Veldhuis, S. A.; Boix, P. P.; Yantara, N.; Li, M.; Sum, T. C.; Mathews, N.; Mhaisalkar, S. G. Perovskite materials for light-emitting diodes and lasers. *Adv. Mater.* **2016**, *28*, 6804-6834.

(11) Kovalenko, M. V.; Protesescu, L.; Bodnarchuk, M. I. Properties and potential optoelectronic applications of lead halide perovskite nanocrystals. *Science* **2017**, *358*, 745-750.

(12) Chen, J. Perovskite quantum dot lasers. *Infomat* **2019**, *2*, 170-183.

(13) De Giorgi, M. L.; Anni, M. Amplified spontaneous emission and lasing in lead halide perovskites: state of the art and perspectives. *Appl. Sci.* **2019**, *9*, 4591.

(14) Xu, J.; Li, X.; Xiong, J.; Yuan, C.; Semin, S.; Rasing, T.; Bu, X. H. Halide perovskites for nonlinear optics. *Adv. Mater.* **2019**, *32*, 1806736.

(15) Dong, Y.; Zhang, Y.; Li, X.; Feng, Y.; Zhang, H.; Xu, J. Chiral perovskites: promising materials toward next-generation optoelectronics. *Small* **2019**, *15*, 39.

(16) Brennan, M. C.; Forde, A.; Zhukovskyi, M.; Baublis, A. J.; Morozov, Y. V.; Zhang, S.; Zhang, Z.; Kilin, D. S.; Kuno, M. Universal size-dependent stokes shifts in lead halide perovskite nanocrystals. *J. Phys. Chem. Lett.* **2020**, *11*, 4937–4944.

(17) Zhu, H.; Fu, Y.; Meng, F.; Wu, X.; Gong, Z.; Ding, Q.; Gustafsson, M. V.; Trinh, M. T.; Jin, S.; Zhu, X. Y. Lead halide perovskite nanowire lasers with low lasing thresholds and high quality factors. *Nat. Mater.* **2015**, *14*, 636–642.

(18) Qin, C.; Sandanayaka, A. S. D.; Zhao, C.; Matsushima, T.; Zhang, D.; Fujihara, T.; Adachi, C. Stable room-temperature continuous-wave lasing in quasi-2D perovskite

films, *Nature*. **2020**, *585*, 53–57.

(19) Yakunin, S.; Protesescu, L.; Krieg, F.; Bodnarchuk, M. I.; Nedelcu, G.; Humer, M.; De Luca, G.; Fiebig, M.; Heiss, W.; Kovalenko, M. V. Low-threshold amplified spontaneous emission and lasing from colloidal nanocrystals of caesium lead halide perovskites. *Nat. Commun.* **2015**, *6*, 8056.

(20) Wu, K.; Park, Y. S.; Lim, J.; Klimov, V. I. Towards zero-threshold optical gain using charged semiconductor quantum dots. *Nat. Nanotechnol.* **2017**, *12*, 1140-1147.

(21) Zhou, C.; Zhong, Y.; Dong, H.; Zheng, W.; Tan, J.; Jie, Q.; Pan, A.; Zhang, L; Xie, W. Cooperative excitonic quantum ensemble in perovskite-assembly superlattice microcavities. *Nat. Commun.* **2020**, *11*, 329.

(22) Fu, J.; Xu, Q.; Han, G.; Wu, B.; Huan, C. H. A.; Leek, M. L.; Sum, T. C. Hot carrier cooling mechanisms in halide perovskites. *Nat. Commun.* **2017**, *8*, 1300.

(23) Liu, Q.; Wang, Y.; Sui, N.; Wang, Y.; Chi, X.; Wang, Q.; Chen, Y.; Ji, W.; Zou, L.; Zhang, H. Exciton relaxation dynamics in photo-excited $CsPbI_3$ perovskite nanocrystals. *Sci Rep.* **2016**, *6*, 29442.

(24) Jana, A.; Mittal, M.; Singla, A.; Sapra, S. Solvent-free, mechanochemical syntheses of bulk trihalide perovskites and their nanoparticles. *Chem. Commun.* **2017**, *53*, 3046–3049.

(25) Krieg, F.; Ochsenbein, S. T.; Yakunin, S.; ten Brinck, S.; Aellen, P.; Süess, A.; Clerc, B.; Guggisberg, D.; Nazarenko, O.; Shynkarenko, Y.; Kumar, S.; Shih, I. Infante, C. J.; Kovalenko, M. V. Colloidal $CsPbX_3$ (X = Cl, Br, I) nanocrystals 2.0: zwitterionic capping ligands for improved durability and stability. *ACS Energy Lett.* **2018**, *3*, 641–646.

(26) Rainò, G.; Becker, M. A.; Bodnarchuk, M. I.; Mahrt, R. F.; Kovalenko, M. V.; Stöferle, T. Superfluorescence from lead halide perovskite quantum dot superlattices. *Nature* **2018**, *563*, 671–675.

# Supporting Information

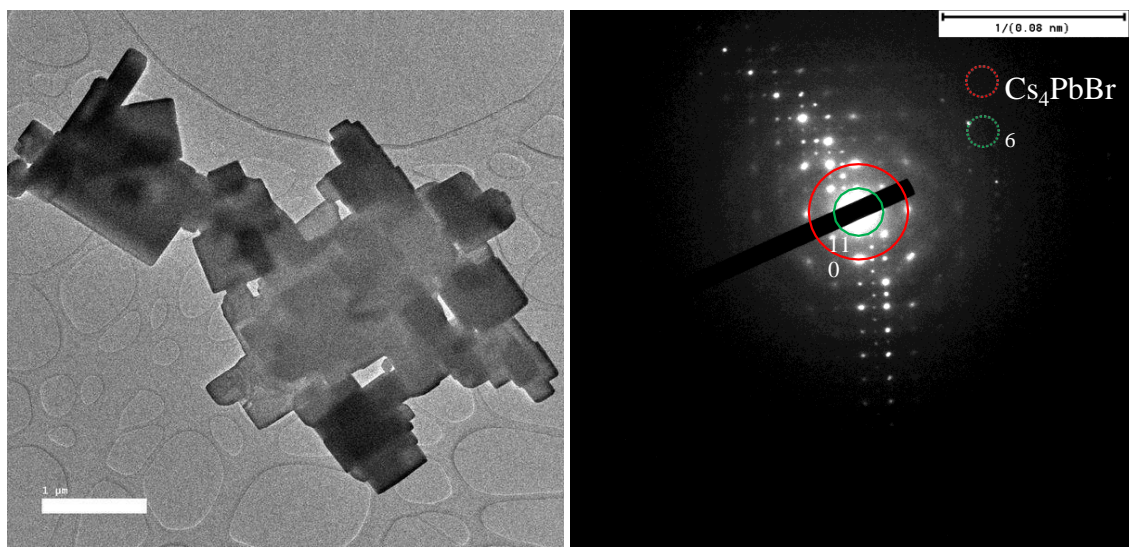

**Fig. S1| TEM image (left) of CsPbBr$_3$ nanocrystals with cuboid symmetry**. The thickness of the sample does not allow fringes of the lattice to be visible but the Selected Area Electron Diffraction (SAED) pattern (right) of the sample corroborates the presence of CsPbBr$_3$ nanocrystal islands encapsulated in Cs$_4$PbBr$_6$ microcrystals with rhombic symmetry.

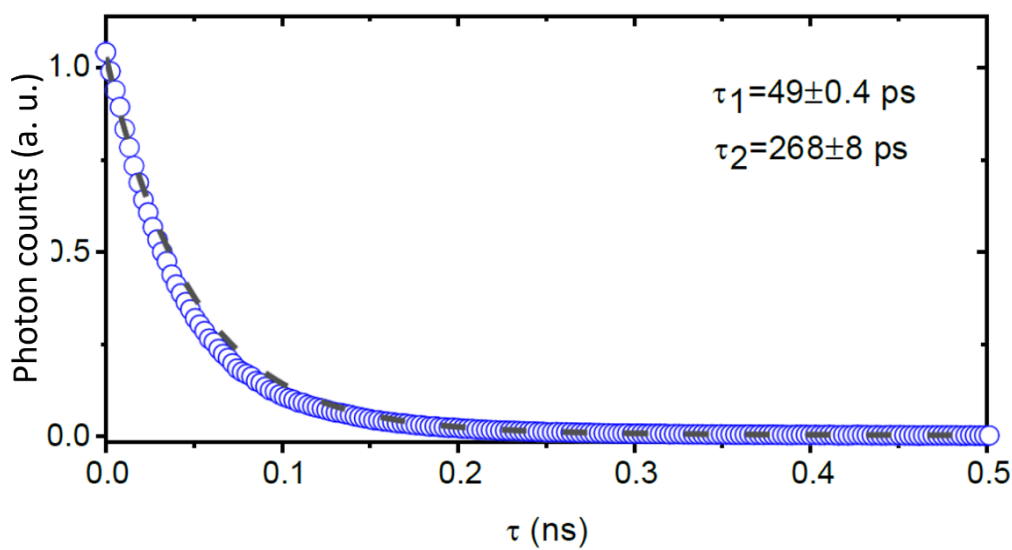

**Fig. S2 | Time-resolved PL decay of CsPbBr$_3$ nanocrystal pumped quasi-resonantly**.

The decay fit was produced using *PicoQuant EasyTau* software to extract the time constants for the fast signal (49 ps) due to stimulated emission and the delayed contribution (268 ps) from spontaneous emission convolved with the instrument response function. The spontaneous lifetime corroborates a weak confinement regime in nanocrystals where the Bohr radius of the exciton is ~10 nm.

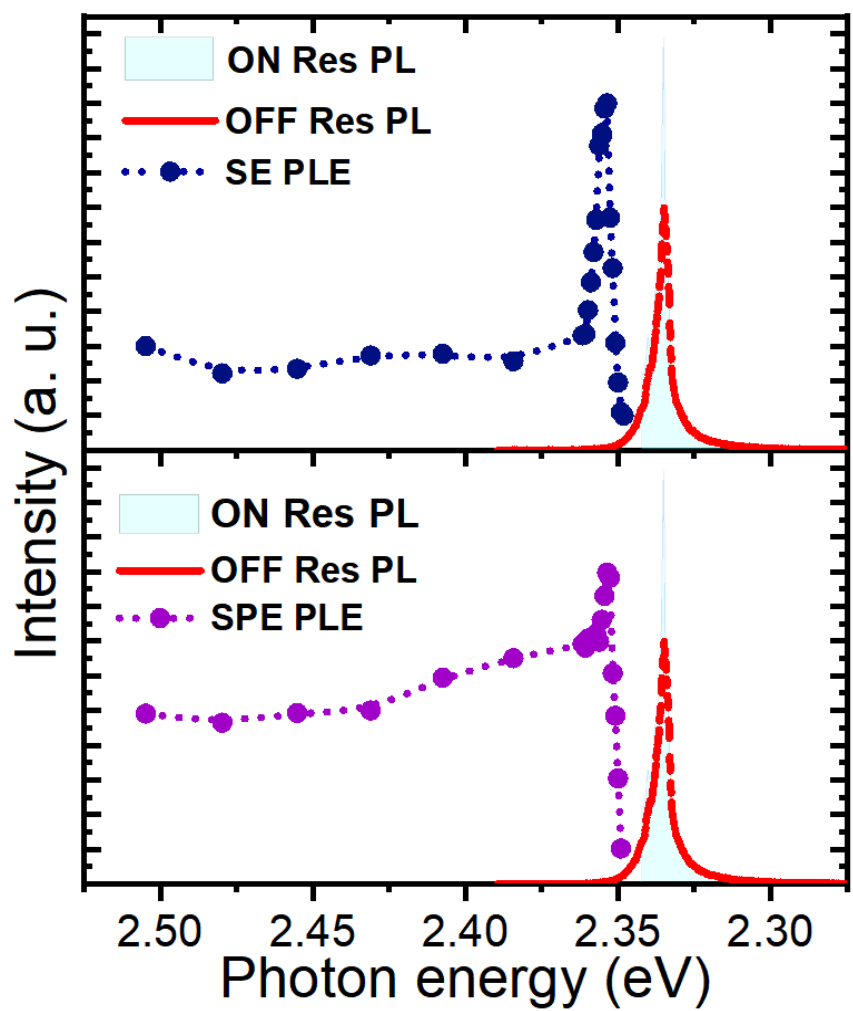

**Fig. S3 | PLE and PL plot of on – and off resonant.**
PLE plot for SE (dark blue dotted line) and SPE (violet dotted line) signals respectively. The red profile shows the PL emission from non-resonant pumping and the light blue shadow indicates that obtained from near-resonant excitation. The SE PLE peak is Stoke-shifted by 18 meV with respect to the SE PL peak, and suggests small losses due to vibrational relaxation.

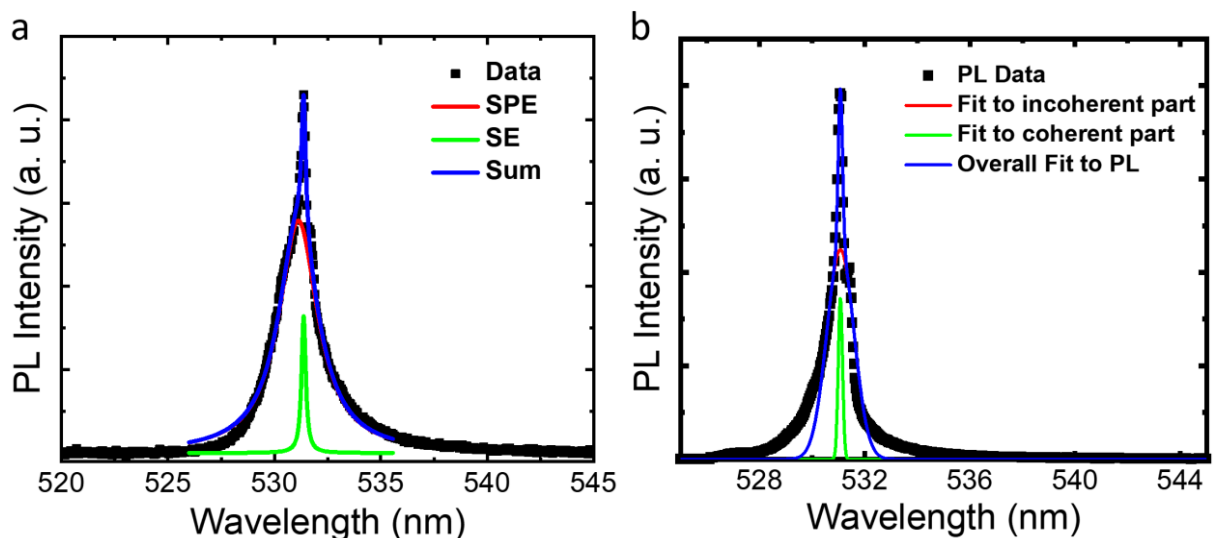

**Fig. S4| PL fitting.**
**(A)** Fitting of off-resonant PL emission using Gaussian function measured at 25 mW pump power. **(B)** Fitting of the coherent and incoherent components of the CsPbBr$_3$ nanocrystal emission signal from the time integrated PL spectrum.

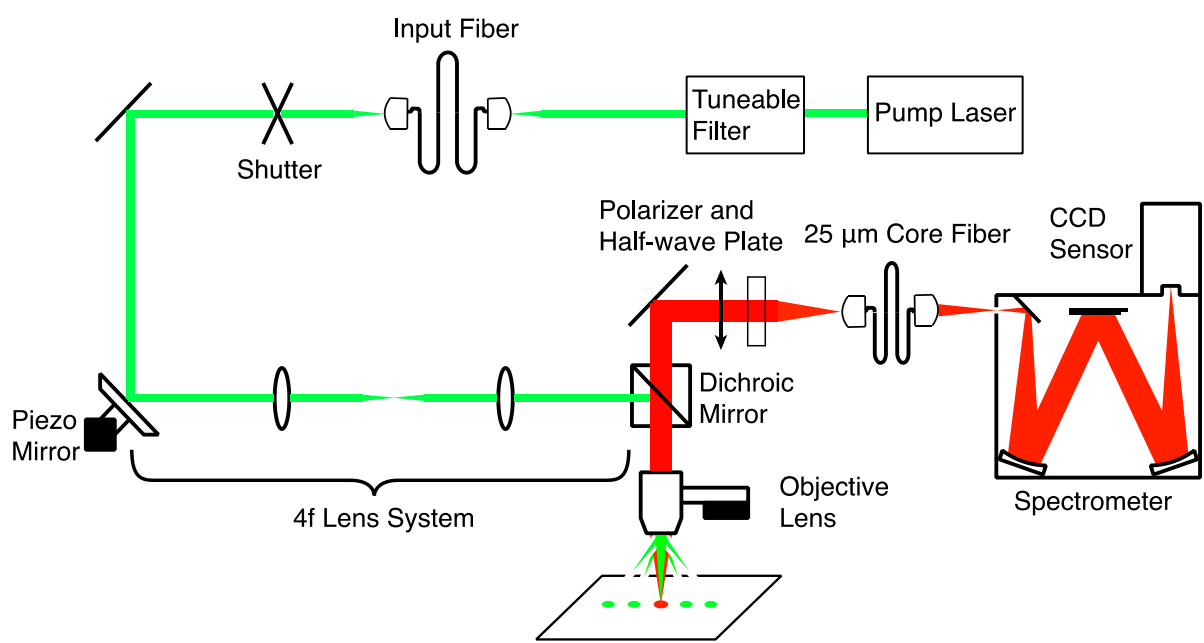

**Fig. S5| Diagram of apparatus.**
Diagram of the telecentric optical system used to excite the perovskite microcrystals and the confocal collection system used to gather the luminescence. The objective lens is mounted on a piezoelectric xyz and can be scanned independently of the excitation. Polarizer and half wave plate were used together to analyze the emission and remove any influence from the polarization dependence of the spectrograph and CCD.